# Lock-in detection using a cryogenic low noise current preamplifier for the readout of resistive bolometers

D. Yvon[1,*], V. Sushkov[2,#], R. Bernard[2], J.L. Bret[3], B. Cahan[2], O. Cloue[4], O. Maillard[2], B. Mazeau[4], J.P. Passerieux[2], B. Paul[2], C. Veyssiere[4].

[1]CEA, Centre d'Etude de Saclay, DAPNIA, Service de Physique des Particules, bat 141, F-91191 Gif sur Yvette Cedex.

[2]CEA, Centre d'Etude de Saclay, DAPNIA, Service d'Electronique d'Informatique, bat 141, F-91191 Gif sur Yvette Cedex.

[3]CNRS-CRTBT, 25 avenue des Martyrs, BP 166, F-38042 Grenoble Cedex 9

[4]CEA, Centre d'Etude de Saclay, DAPNIA, Service d'Instrumentation Générale, bat 472, F-91191 Gif sur Yvette Cedex.

[*]Corresponding author, electronic mail address: yvon@hep.saclay.cea.fr, Tel: (33) 0169083911, Fax: (33) 0169086428

[#]Now at Lawrence Berkeley Laboratory, Berkeley, CA, USA.

PACS: 85 (Electronics), 07.57K (Detector, bolometer), 07.20.M (Cryogenics instrumentation)

We implemented a low noise current preamplifier for the readout of resistive bolometers. We tested the apparatus on thermometer resistances ranging from 10 $\Omega$ to 500 M$\Omega$. The use of current preamplifier overcomes constraints introduced by the readout time constant due to the thermometer resistance and the input capacitance. Using cold JFETs, this preamplifier board is shown to have very low noise: the Johnson noise of the source resistor (1 fA/Hz$^{1/2}$) dominated in our noise measurements.

We also implemented a lock-in chain using this preamplifier. Because of fast risetime, compensation of the phase shift may be unnecessary. If implemented, no tuning is necessary when the sensor impedance changes. Transients are very short, and thus low-passing or sampling of the signal is simplified. In case of spurious noise, the modulation frequency can be chosen in a much wider frequency range, without requiring a new calibration of the apparatus.

## Introduction

Low temperature bolometers address detection problems that require high sensitivity, high energy resolution and/or low thresholds: X-ray and far infrared spectroscopy, double beta decay, Dark Matter particle searches. To reduce thermal noise they must be operated at temperatures generally in the 10-300 milliKelvin range. Their performance critically depends on the quality of the thermometer and on the readout electronics and wiring of the apparatus. This work relies on the experience acquired by the detector community [1, 2]. In ref. [3] we discussed different preamplifiers configurations. This paper describes the implementation and test results of the current preamplifier, and the design of a lock-in detection chain using it.

## Cryogenic constraints:

We assume in the following that the thermometers used are resistive, of the order of 1 MΩ resistance, or more. The signal is a variation of the thermometer resistance $\Delta R_b$.

In this context, JFET, MOSFET, GaAs MESFET and HEMT are natural candidates for front-end devices. For example, Alessandrello et al. [5] designed a voltage preamplifier using GaAs transistors, working at a temperature of 1K, and dissipating only 54 mW. The main drawback of MOSFET and MESFET is that up to now, their low frequency noise performance is poorer than the commercially available Si JFET, which on the other hand are not able to work at such temperatures. Since we expect detector Johnson noises of the order of few $nV/Hz^{1/2}$ (few $fA/Hz^{1/2}$), we were led to choose JFETs, known to achieve voltage noises of the order of 1 $nV/Hz^{1/2}$ and current noises of the order of 0.1 $fA/Hz^{1/2}$. One could consider placing the full preamplifier at cryogenic temperatures as in [5], but this is expensive because of liquid helium consumption. We decided to run the preamplifier at room temperature, only the JFET being cold. We used JFET power dissipation ranging from 0.5 mW to 10 mW, per channel: the more power in the JFET, the lower the voltage noise. In some cryostats, the JFET can be anchored to the liquid nitrogen shield, where thermal dissipation is not expensive. But most dilution or $^3$He cryostats do not allow this, and the JFET has to be placed at the 4K stage. Then the user has to make a compromise between the helium boiling rate and voltage noise in the JFET. Unfortunately, silicon JFETs do not work at 4 K and to minimise voltage and current noise, the JFET has to be cooled to a temperature between 130 K and 180 K. Thus we had to thermally isolate the JFET.

The temperature increase of the bolometer (due to particle interaction or Infrared incoming power) induces a variation of the thermometer impedance. Signals may be slow: from 1 second to few μs. We wish then to use a DC-coupled preamplifier. When powered, the JFET self heats to a temperature around 150 K. Gate-to-source JFET voltage depends on the temperature. To prevent the amplification of the offset and drift voltages one may use a differential preamplifier [6]. This option is expensive in a cryogenic environment, because it requires four wires per channel to bias the JFETs of the first stage, doubles the power dissipation at 4 K, and increases the white voltage noise by a factor of $\sqrt{2}$. We choose to use single JFET, with the drawback that it is more sensitive to pickup noise and a biasing regulation of the JFET is necessary. This JFET biasing regulation will produce a low-frequency high-pass filter behaviour in the preamplifier.

The coldest stage of most low temperature cryostats cannot handle power dissipations larger than 1 μW. Thus the readout JFET cannot be located next to the detector. In our apparatus front end JFETs are placed at 20 to 40 cm from the detector. Due to the thermometer high impedance, microphonics and electromagnetic noise pickup as well as the capacitance of the wiring then become crucial problems, and great care has to be taken to optimise the wiring connections. We decided to design low capacitance, low-noise cryogenic coaxial cables. The performance obtained, turned out to be quite satisfactory.

To summarise the configuration, we chose to design a low-noise, wide-bandwidth, DC-coupled JFET preamplifier. Active regulation of the current flowing through the JFET suppresses frequencies below 0.1 Hz. The power applied to the JFET ranges from 0.5 mW to 10 mW. It is thermally anchored to the 4.2K liquid helium or 77 K liquid nitrogen bath, depending on the cryostat schematic, through a thermal impedance computed to achieve a temperature close to 150 K.

## Preamplifier design

From the electronics point of view, the goal was to design a low-noise preamplifier to obtain the best threshold and resolution performances. So as to allow detector characterisation and optimisation, the readout had to induce minimal distortion of the thermal signal of the bolometer. But large thermometer impedance and stray capacitance work as a low-pass filter. In reference [4], we implemented an "active bootstrap circuit" to overcome this problem. In this paper we expand on what Silver et al. [7] had previously used to readout fast microcalorimeters.

### Schematic

Figure 1 shows the simplified schematic of the current preamplifier configuration. A heat pulse induces a resistance change of the thermometer. A biasing voltage $V_p$ is applied to the thermometer. The feedback around the operational amplifier forces the voltage at its inverting input to be zero. Consequently the input capacitance $C_s$ does not charge. Thus the signal is not low-passed with the time constant $R_b C_s$. In the limit of small thermometer impedance change,

$$\Delta V_{out} = V_p \frac{\Delta R_b}{R_b^2} \frac{R_f}{1 + j\omega R_f C_f} \qquad \text{equation (1)}$$

where the signal is the variation of the bolometer resistance $\Delta R_b$, $\omega$ is the angular frequency, $R_b$ and $R_f$ are the bolometer and feedback resistances, $C_b$, $C_f$ are the bolometer and the feedback stray capacitances, and $C_s$ the input plus wiring stray capacitance of the preamplifier. To achieve a current preamplifier in the context presented in the introduction, we designed a board whose schematic is given in figure 2.

The electronic circuit runs at room temperature outside the cryostat, except for the JFET, staying at 77K or 4 K, and the bolometer and feedback resistor placed on the cold plate of the cryostat. The JFET input capacitance is 12 pF and the wiring stray capacitance depends of the cryostat implementation.

### Preamplifier loop

The preamplifier loop (figure 2) includes two circuits. In the first a very fast common base current amplifier of gain 5 (R6/R7) amplifies the small signal current coming from the JFET. This amplifier raises the current above the input current noise of the following AD844. The AD844 further amplifies the current with an open loop transresistance gain of 3 MΩ yielding the preamplifier output.

FET biasing circuit

From the schematic, when no voltage bias is applied, we expect the output of the preamplifier to be close to the gate voltage of transistor J1, since the gate current of the cold FET is negligible. To null the output voltage, we have to adjust the JFET source voltage, to ensure the proper DC conditions. This voltage depends of the FET temperature and an active correction of offset is necessary to prevent the output from saturating. This is the purpose of the OPA111 based integrator circuit. Assuming no input voltage from "Polar DC" (a lock-in symmetrical square waveform is used), the integrator integrates the output voltage of the preamplifier. Through R22, the OP111 output drives the JFET source voltage and cancels the offset drifts. The time constant of the integrator divided by the looped gain (~0.09*(Rf+Rb)/Rb) has to be much slower than the pulse time. We chose the time constant as 1s to ensure that a preamplifier is active one minute after being powered. The biasing voltage is then low-pass filtered to suppress noise from OPA 111. At frequency below 20 Hz, the OPA111 1/f noise may be dominating the noise budget. If such a use is planned, the OPA111 should be replaced by an AD743. Then the R18 Johnson noise will be the dominating voltage noise (only below 7 Hz) if the JFET is cooled down to 150K, with a noise referred to the input of the preamplifier of 1.1 nV/sqrt(Hz).

Thermometer biasing circuits

Two common biasing conditions had to be implemented: when used to measure slow signals AC biasing and lock-in detection is often optimal, when used with fast bolometers for particle interactions DC biasing is preferred. The difficulty comes from the fact that modulation is applied to the lowest noise stage of the preamplifier so the modulation noise has to be lower than preamplifier noise. Good oscillators generate signals of volts amplitude with noise level lower than 100 nV/Hz$^{1/2}$. This has to be divided by at least a factor 100 to meet our requirements. This is the purpose of the R28 and R29 resistor divider. The carrier signal then feeds directly to the thermometer. We emphasise that the JFET biasing system requires the modulation frequency to be above 20 Hz.

Applying a DC voltage to bias the thermistor is somehow incompatible with the FET biasing circuit. If a constant voltage V is applied through the AC biasing circuit of the bolometer, the FET biasing circuit changes the FET source voltage to recover ground at the preamplifier output. Thus the FET gate potential rises to $V_g = \frac{VR_f}{R_f + R_b}$, inducing an effective polarisation voltage on the bolometer $V_p = V\frac{R_b}{R_f + R_b}$. Typical operating biasing potentials of thermometer are in the 10 mV range, and the resistor ratios Rf/Rb are typically 10. This induces gate potentials of a few 100 mV, and sets two potential problems. First when the gate wire is no longer grounded, microphonics interference happens through the varying capacitance between the gate wire and its grounded shielding. Second, because of low power cryogenic requirements, the FET is operated with a drain voltage below 3 volts. If Rf/Rb≥100, changing the source and gate potential by 1V may drive it out of the saturation zone (where its has lower gain and its white noise increases). To prevent this, we implemented an additional feature for DC biasing. We apply a DC Voltage $V_{DC}$ at the input of the DC polarisation AMP03 chip and the AC biasing circuit is grounded. This changes the reference voltage of the JFET biasing, the preamplifier output rises to $V_{DC}$ and the FET gate potential rises to $V_g = \frac{V_{DC}R_b}{R_f + R_b} = V_p$ much lower than the previous configuration. The drawback is that the output offset has to be removed before further amplification of the signal, but we assumed our signals were relatively fast in this detection mode. When we wish to substract the DC component we remove the short-circuit around the C5 decoupling capacitor .

Post amplification and power supply

The preamplifier, being not symmetric, has poor common mode noise rejection properties. Very careful power supply noise filtering is required. We use low noise linear power supplies and the classical low pass filter around 2N2905, BC560C and 2N2219A, MPSA18 transistors to power the card. To prevent "motorboarting" and to further filter the power supply, we added large value capacitors on the power supply lines of the preamplifier.

The AD829 operational amplifier and the LM6321 driver provide post amplification and allow driving post stage electronic cards. To reduce pick up in the incoming and outgoing lines, we used shielded twisted pair cables, differential amplifiers (AMP03) at the voltage inputs and we symmetrised the impedance of the two leads using 1kΩ resistors keeping this design philosophy in the lock-in boards design.

**Test results**

Noise measurements

Noise measurements were performed in two conditions. In the first one we used a 300 K JFET, and liquid nitrogen cooled high value feedback resistors. We operated our apparatus with feedback resistors up to 10 GΩ, the bolometer not connected. The bandwidth of the preamplifier for the feedback resistor Johnson noise was limited to $\omega_f = 1/(R_f C_f)$. The total noise measured when referred to the input of the amplifier (figure 3) accounted for a current noise as low as $10^{-15}$ A/Hz$^{1/2}$, dominated by the Johnson noise of the feedback resistor. In the second test setup, noise measurements were performed within a 300 mK cryostat. As described in the introduction, the JFET ran at 150 K in a 77 K environment. A 10 MΩ feedback resistor was placed on the cold plate at 320 mK, and again no bolometer was connected to maximise the input impedance. However we used regular cryogenic coaxial wiring, because we cannot afford the heat load of the available low noise coaxial wires. Not surprisingly, the observed spectrum displays strong microphonics and 50 Hz coming from the main lines. Nevertheless, on the plateau we measure $i_n \sim 1.8$ fA/Hz$^{1/2}$, corresponding to the predicted value of resistor Johnson noise. Previous measurements, using a 50 MΩ resistor placed at 20 mK on the cold plate of a dilution fridge, showed the cold JFET current noise is as low as 0.2 fA/Hz$^{1/2}$.

Bandwidth measurements

For this test, we placed the cold circuits in the 300 mK cryostat. A 1 MΩ resistor was readout by a NJ132L Interfet JFET, and placed in series with a $R_f = 10$ MΩ MiniSystem cold resistor [13]. Then we plugged a regular voltage amplifier [4] and measured a bandwidth of 2.05 kHz, corresponding to an input capacitance of 80 pF. Plugging the above current preamplifier, we measured a bandwidth of 45 kHz, much larger than the previous one as expected. The corresponding stray capacitance on the feedback resistor is $C_f = 0.36$ pF, in accordance with the MiniSystem specifications of 0.2 pF, plus an additional capacitance ascribed to the wiring of the surface mount resistor chip.

Stability

The stability of this preamplifier has been tested with bolometer resistance $R_b$ ranging from 10 Ω to infinity (open circuit). For low $R_b$ values, the wide band feedback through $R_f$ and $R_b$ does not limit the gain. The slow feedback loop gain is large and low frequency oscillations may happen as the feedback circuit has two poles, the integrator and the low pass filter. To prevent this predictable oscillation, we placed a 30 kΩ resistor (R 18) in series with the integrator capacitor. This cancels out the pole of the filter. Then we experienced that the resistor value $R_5$ had to be larger than 1.5 kΩ, to ensure stability. Due to massive filtering, (10 Ω decoupling resistors, and 10 μF decoupling capacitors have been omitted on the schematic for clarity) the impedance of the power supply range from 1 Ohm (second stage) up to ten Ohms (preamplifier loop). When the first stage gain is not limited anymore by the feedback loop, the load on the AD844 power supply is large enough to induce reaction on the preamplifier and low frequency oscillations. The R5 value (2.5 kΩ) was chosen as a compromise with the second stage amplifier noise ( 6.5 nV/sqrt(Hz) ) and stability.

Practical limitations.

Technological limitations to the rise-time appear. The main one comes from the availability of low temperature feedback resistors. The rise time is set by the feedback resistor pole $\omega_f = 1/(R_f C_f)$. Most high ohmic value resistors diverge to very high resistance when placed at temperatures lower than 4K. Using the MSI 10 MΩ resistors[13] available on the market, ($C_f = 0.2$ pF) we expect a rise time of 4.4 µs, independent of the thermometer impedance.

It must be pointed out that the thermal relaxation time of the bolometer is slower using the current amplifier configuration than a looped or unlooped voltage amplifier. Also, under extremely strong biasing, the detector and preamplifier chain could become thermally unstable. This behaviour was never observed by us as expected from simple calculations (see for example in reference [3]). If one of these two points turn out to be a problem, the same preamplifier can be used as a good voltage amplifier with feedback looped on the bolometer [7]. This readout configuration allows the advantages of fast risetime and avoids electrothermal instability and enlarged thermal relaxation time. A comprehensive discussion of this issue is presented in [3].

## **Lock-in design using the current amplifier**

When the thermal signal is slower than a few tens of millisecond, classical readout electronics includes a lock-in amplifier to extract the signal from 1/f preamplifier noise and out of band microphonics. We designed test boards to demonstrate the performance and convenience of a lock–in design using a current amplifier. So as to simplify design and debugging we placed the preamplifier, oscillator and modulator/demodulator on three distinct boards. To prevent pick up, necessary connections were achieved through shielded twisted pair cables. This choice necessitated the use of numerous differential receivers and driver chips. In future implementations of this circuit, these could be removed using a single board design. The test boards were made of a few modules, whose schematics are drawn at figure 4.

Oscillator

A pulse generator provides an adequate frequency reference that drives the modulator and the time-delayed carrier mono-polar square waves. These are used to trigger the modulator and demodulator AD630 chips.

Modulator

The modulator provides a square wave signal of chosen amplitude. So as not to degrade the resolution three main features must be achieved. The noise of the square wave has to be smaller than 100 nV/Hz$^{1/2}$. Chips available on the market meet this requirement easily. More difficult is to achieve an amplitude stability of few $10^{-6}$/°C for the digitally chosen signal amplitude. This uncommon feature can be achieved on square wave signals using high performance DAC resistor bridges and a good voltage reference. The LM299 voltage reference output is chopped by the AD630 to produce a fixed amplitude stable symmetric square wave. This square wave enters the reference voltage input of a DAC 7541A and is divided by the DAC high stability resistor's bridge and AD743 operational amplifier. The C3 capacitor allows smoothing of ringing due to stray capacitors in the DAC. The output provides the square wave biasing of the thermometer. Every part or chip relevant for the amplitude stability was chosen to have a temperature coefficient smaller than $10^{-6}$/°C.

Demodulator

The preamplifier provides a detected square wave with minimal distortion (figure 5a): the phase shift (time delay) is small and does not depend on the sensor impedance, but only on the feedback resistor. Compensation of time delay might be unnecessary. If implemented, it will reduce transient amplitudes and the delay adjustment is fixed: no tuning is necessary when sensor impedance changes.

The demodulator block has to meet the same amplitude stability requirement. SSM2141 chips provide adequate temperature coefficients. Then the time-delayed oscillator output triggers the AD630 chip that rectifies the square wave. The resulting signal is close to a constant voltage (figure 5b) with a fast spike down to ground.

Outputs

The DC value of AD630 is directly related to the thermometer impedance, and may be useful data. Thus we implemented a "slow" output after a 2 Hz low-pass filter.

Most of the time however we are interested in faster very small signals. Thus to prepare the "fast" output signal, we subtract the signal DC value, and further amplify the signal using an OP27A. To generate the substracted voltage, we use again a DAC7541A chip, and the same voltage reference as for the modulation (same board). If this voltage reference drifts, then it generates no DC signal on the fast output. After baseline subtraction, a temperature coefficient of $10^{-4}/°C$ is adequate. Commercial products commonly achieve this and temperature stability of electronic parts is not a concern anymore.

Baseline subtraction and filtering issues.

Baseline subtraction induces another noise contribution. To reject modulation transients, the lock -in output will have to be lowpassed. The low-passed-signal DC value is offset from the "true" value by the average of the integral of transients, that depends on modulation frequency. This implies that the gain of the lock-in loop depends somewhat on the modulation frequency. If the modulation frequency is unstable, the induced offset and gain change. This additional noise voltage is less than $e_{offset}$ given by:

$$e_{offset} = \frac{ATf}{G}\left(\frac{\Delta f}{f}\right)$$

where $e_{offset}$ is the equivalent noise contribution (V/Hz$^{1/2}$) at the input of the preamplifier, A and T are the amplitude and the duration of transients, G is the gain of the electronic chain before subtraction, and f is the modulation frequency. For typical values (A= 10 V, T= 10μs, G=5.10$^3$, f=1kHz), a frequency stability of $10^{-4}$ is enough to ensure negligible noise contribution.

After demodulation and offset subtraction, the output signal has to be low-pass filtered in order to get rid of the short transients (Figure 5b). This is a relatively easy task, since the transients are of high frequency (more than 50 kHz) far above the harmonics of regular modulating frequencies. But unlike in most lock-in designs in order to remove the high frequency preamplifier white noise, the filter has to be at least of second order [3].

## Test results

We measured noise spectra of the full lock-in chain. All measurements showed that the noise of the lock-in chain is dominated by the preamplifier contributions. We then used the above preamplifier and lock-in to study two bolometers, one constructed by Infrared Labs [11] and one fast bolometer model useful for particle detectors.

As a validation tool for these developments we constructed a 300 mK cryostat and instrumented it with four channel cold JFET readout electronics. The cryostat is a Infrared Space Observatory (ISO) test cryostat [8] consisting of liquid nitrogen cooled first stage, on which the JFET and wiring are thermally anchored and a pumped $^4$He 1.8 K cold plate, where all the wiring is thermalised. Finally, there is a single shot pumped $^3$He cryocooler from L. Duband[9], thermally anchored to a 100 cm$^2$ cold plate. This apparatus achieved a temperature of 315mK, limited by the thermal load from the 25 cryogenic coaxial cables used for the readout electronics. An AVS46 resistor bridge is used to monitor the cryostat temperature and a Gould 1604 digital oscilloscope to register the signals. The following measurements have been made using the above presented current preamplifier and when necessary the lock in chain.

### Infrared bolometer

We first measured the IV curves of the Infrared Labs bolometer [11] kindly provided by J-M. Lamarre. Figure 6 plots the results of the measurements, presented as the "blind" bolometer impedance (no infrared light power impings the bolometer) versus the biasing thermal power load. This allowed us to determine the thermometer law R(T) and the thermal conductivity to the cold bath. We find $R(T) = R_0 \left(\dfrac{T}{1K}\right)^{-8.3}$, with $R_0$= 27.9 k$\Omega$, and a thermal conductance to cold bath at 373 mK well fitted by a power law: P($\Delta$T) = 2.9 10$^{-9}$ ($\Delta$T)$^2$ , P in Watts and $\Delta$T in Kelvin. This shows that the heat leak is complex, not dominated by a Kapitza resistance or a metal-like conductivity.

Current amplifier technology allowed us to use a modulation frequency of 20 Hz for these measurements on thermometer resistances up to 450 MΩ and a total input capacitance of 82 pF, with no loss in signal amplitude.

## Pulses data acquisition

We use the current preamplifier to readout a fast bolometer model, made of a 8x12x0.3mm$^3$ Sapphire test plate onto which an NbSi thin film thermometer [10] is evaporated. Thermal connections were made with gold wires, wire-bonded on the Sapphire plate. An $^{241}$Am alpha source induced the detected heat pulses.

Figure 7 plots typical alpha events obtained at a temperature of 350 mK. The pulse shape displays two different regimes. A fast part, and a slow relaxation. As stated in [10,12], the fast part is due to conversion of out of equilibrium phonons in the thin film thermometer: the thermometer overheats relative to the Sapphire substrate. Within few tens of microseconds, the thermometer temperature relaxes to the substrate temperature, and then the whole device temperature relaxes through the heat leak. Had we been using a regular voltage amplifier the fast signal part would have been low-passed filtered by the input capacitance and thermometer resistance (80 pF and 300 kΩ) of this device.

Historically, this effect remained unknown to us until it was found [12] on a low resistance thermometer (few kΩ).

## **Conclusion**

We implemented a current preamplifier for the readout of resistive bolometers. We tested the apparatus on thermometer resistances ranging from 1 kΩ to 500 MΩ. We showed that the current amplifier allows much flexibility and convenience in the design.

The current preamplifier overcomes constraints introduced by the readout time constant due to the thermometer resistance and the input capacitance. It allows the study of pulse shapes and detector behaviour that would be unnoticed using a voltage amplifier. Using a cold JFET, this preamplifier board is shown to have very low noise: the Johnson noise of the source resistor (1 fA/Hz$^{1/2}$) dominated in our noise measurements.

We also implemented a lock-in chain. We showed that it is very convenient to use the current preamplifier in a lock-in chain, with square wave biasing. Compensation of phase shifts might be unnecessary. If implemented, the adjustment is fixed and no tuning is necessary when sensor impedance changes. Transients are very short, and thus low-passing and sampling of the signal is simplified. In case of spurious noise, the modulation frequency can be chosen in a much wider frequency range, without requiring a new calibration of the apparatus.

## **Acknowledgement**

We are grateful A. Benoit, (CRTBT Grenoble, France), for stimulating discussions, to J-P.Torre (Laboratoire d'aéronomie, Verrière le Buisson, France), L. Dumoulin and P. Garoche (Universite d'Orsay, France) for their advice in designing low noise cryogenic wiring of the cryostat. We want to thank M. Taluro (DESPA, Observatoire de Paris-Meudon) for his advice and help using the ISO cryostat and J. Rich and the Referee for carefully reading proofs of this paper.

## **References**

[1] L. De Carlan et al., "Discrete analog electronics for high-resistivity silicon detectors used for X-rays and γ-rays spectroscopy in whole-body monitoring" Nucl. Inst. Meth. in Phys. Res. A 380 (1996) 371-375

[2] V. Radeka, "Low-noise techniques in detectors", Ann. Rev. Nucl. Part. Sci. 38 (1988) 217-277; N. Madden private discussions.

[3] D. Yvon and V. Sushkov, "Low noise cryogenic electronics: preamplifier configurations with feedback on the bolometer", IEEE Trans. on Nucl. Sci., 47 (2000) 428-437.

[4] D. Yvon et al, "Low noise voltage and charge preamplifiers for phonon and ionization detectors at very low temperature", Nucl. Inst. Meth. in Phys. Res. A 368 (1996) 778-788

[5] A. Alesandrello et al., "Cryogenic voltage–sensitive preamplifier using GaAs MESFETs of low 1/f noise", Nucl. Inst. Meth. Phys. Res. A 295 (1990) 405-410

[6] D.V. Camin et al., Alta Frequenza 56 (1987) 347

[7] Silver E. et al., "High resolution x-ray spectroscopy using germanium microcalorimeters", in "X-Ray, and Gamma-Ray Instrumentation for Astronomy and Atomic Physics", SPIE, vol. 1159 EUV, (1989) 423-432

[8] Observatoire de Paris, Département de Recherche Spatiale DESPA, 5 place Janssen, 92195 Meudon Cedex, France.

[9] L. Duband, CEA-Grenoble, Service des Basses Températures, 17 Av. des Martyrs, F-38054 Grenoble Cedex 9, France.

[10] S. Marnieros et al., "Development of massive bolometers with thin film thermometers using ballistic phonons", Proc. of the VII$^{th}$ International Workshop on Low Temperature Detectors (LTD-7), Max Planck Institute of Physics, Munich, (July 1997) 134-136.

[11] Bolometer #1876, Infrared Laboratories, 1808 East 17$^{th}$ Street, Tucson, AZ 85719–6505, USA.

[12] D. Yvon et al., "Evidence for signal enhancement due to ballistic phonon conversion in NbSi thin film bolometers", Nucl. Inst. Meth. in Phys. Res. A370 (1996) 200-202

[13] Mini-Systems Incorporated, Attleboro, MA 02703, USA

## Figures captions

Figure 1) A simplified schematic of the current preamplifier configuration. $R_b$ and $C_b$ are the bolometer thermometer resistance and capacitance. $R_f$ and $C_f$ are the feedback resistor and capacitance. $C_s$ is the input plus wiring stray capacitance of the preamplifier.

Figure 2) Schematic of the current preamplifier board we implemented. For clarity we omitted the decoupling capacitors and resistors on the power lines of the operational amplifiers, and first stage preamplifier. We delimit with dashed lines the preamplifier loop, the JFET biasing circuit, the thermometer biasing circuit and the second stage amplifier and driver. See text for more details.

Figure 3) Noise spectrum referred to the input, as measured with the current looped preamplifier and $R_b = 1 G\Omega$. The flat noise level is measured at 1.0 fA.Hz$^{-1/2}$. AC Industrial noise pick up in cables at 50 Hz is clearly seen. The bandwidth was limited by the stray capacitance of the 10 G$\Omega$ feedback resistor (250 fF).

Figure 4) Modular schematic of the lock-in chain we developped. The classical functions are implemented: Modulator, preamplifier, demodulator, and baseline substraction. The design is simplified because of the fast risetime of the current looped preamplifier: when the thermometer resistance changes, no adjustement of the phase of the demodulator is necessary. If, due to difficult noise conditions, it is necessary to modulate at a frequency above $\omega_b$, we do not suffer from signal amplitude loss, and therefore, there is no need to recalibrate the detector and electronics chain. The main chips are: AMP03 (differential amplifiers), LM6321 (50 Ohms line drivers), LM299 and MAX672 (voltage references), AD630 (balanced Modulator/Demodulator), AD790 (fast comparator), DAC7541A (high stability DAC). The remaining chips are operational amplifiers and logic operators.

Figure 5) Wave forms as measured at different stages of the lock-in board. The bolometer circuit was simulated with a 1MΩ passive resistor ($R_b$), $C_s$ was set to 100 pF, and $R_F$ to 10 MΩ. The modulation frequency is 2kHz. Figure 5a shows that the modulation square wave is minimaly distorted at the output of the preamplifier. After demodulation and amplification (no offset subtraction requested, figure 5b), the signal is rectified, transients are very fast and thus relatively easy to filter out.

Figure 6) Measured charge curves (resistance versus power) of an Infrared Labs Bolometer placed in and $^3$He single shot cryostat, using this preamplifier board and lock-in. We were able to measure thermometer impedance with power dissipation on the bolometer down to $10^{-16}$ Watt.

Figure 7) Fast pulses registered on digital oscilloscope, from a home-made bolometer model using a Sapphire test plate and an evaporated NbSi thin film thermometer [10]. Two $^{241}$Am alpha pulses are shown. The trace is plotted on the millisecond time scale and displays the thermal relaxation through the heat leak. The inset is a zoom on the fast part of the pulse (time scale of a few hundred of microseconds). Current amplifier technology prevents the low pass filter (due to thermometer impedance and input capacitance) that washes out the fast part of the pulse.

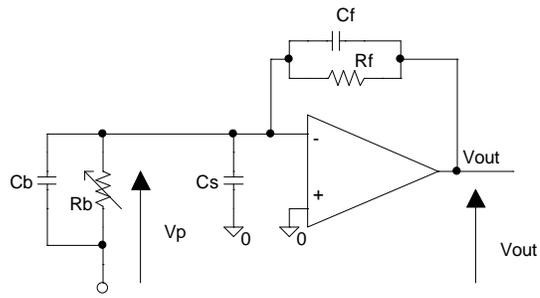

Looped current amplifier

Figure 1

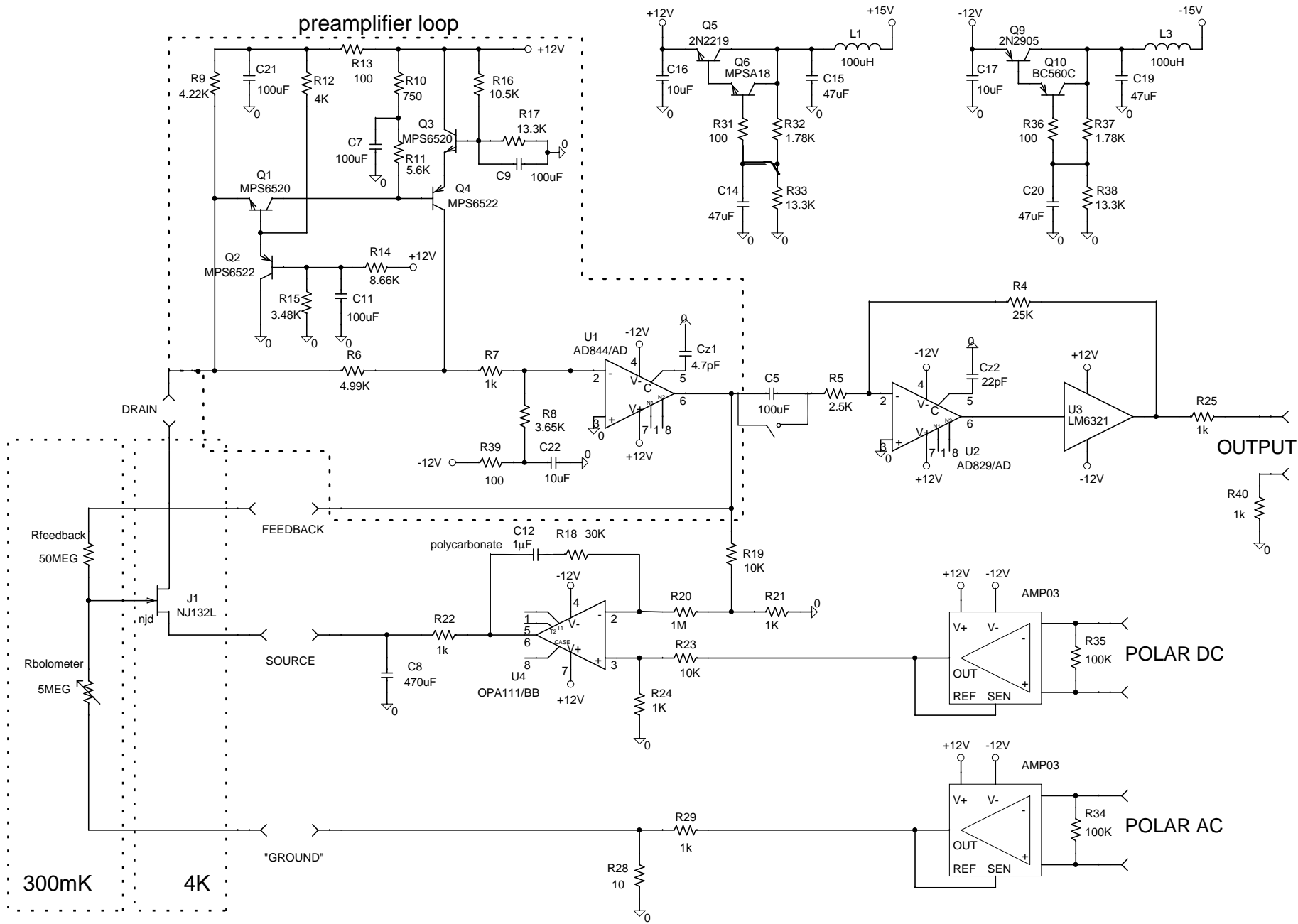

Figure 2

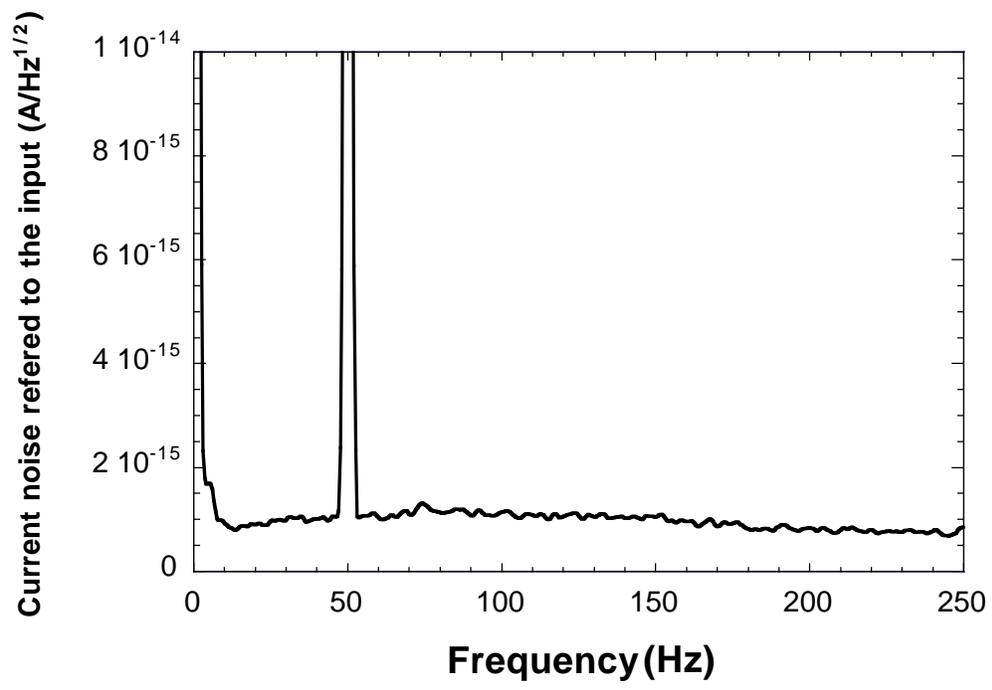

Figure 3

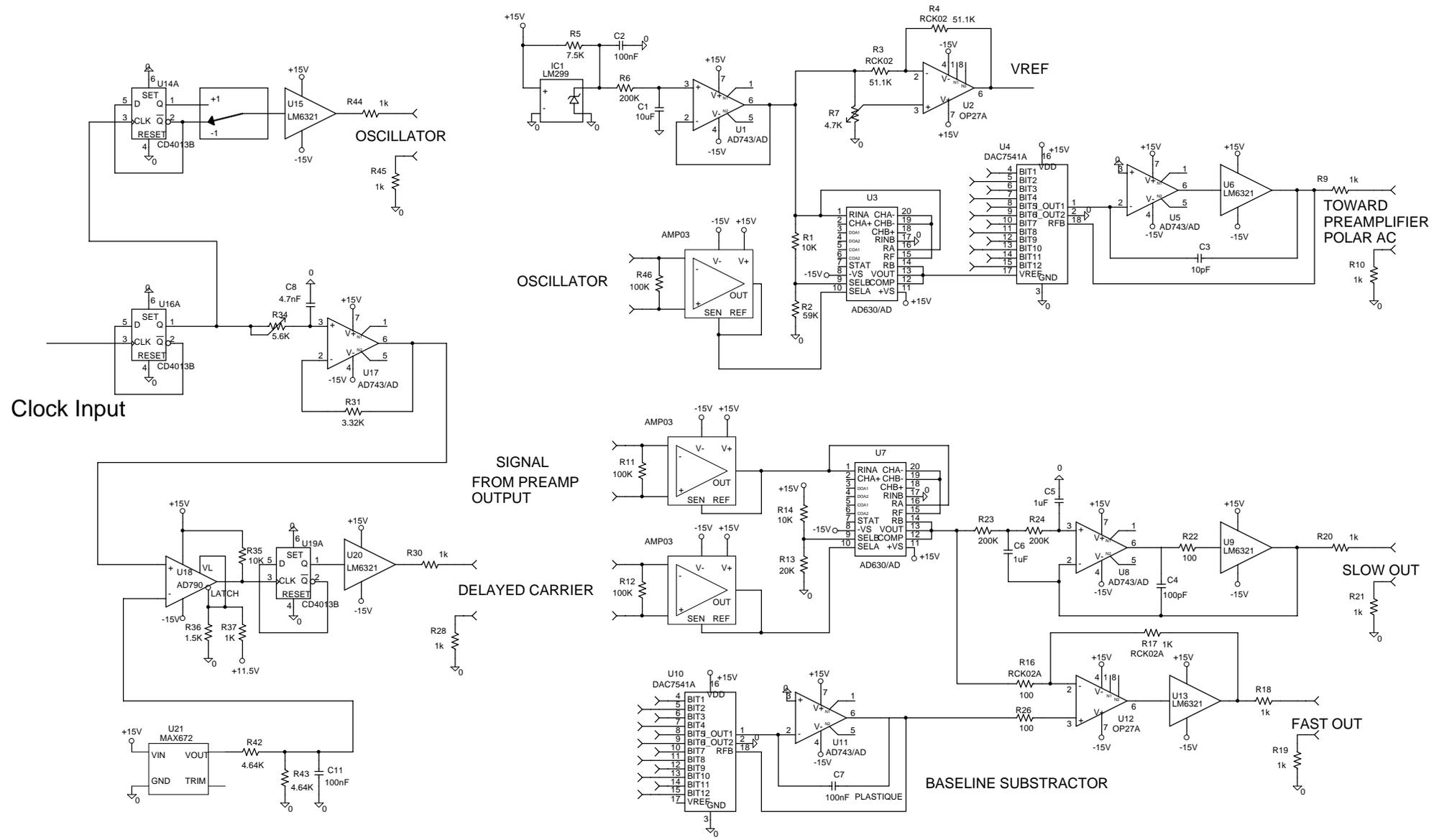

Figure 4

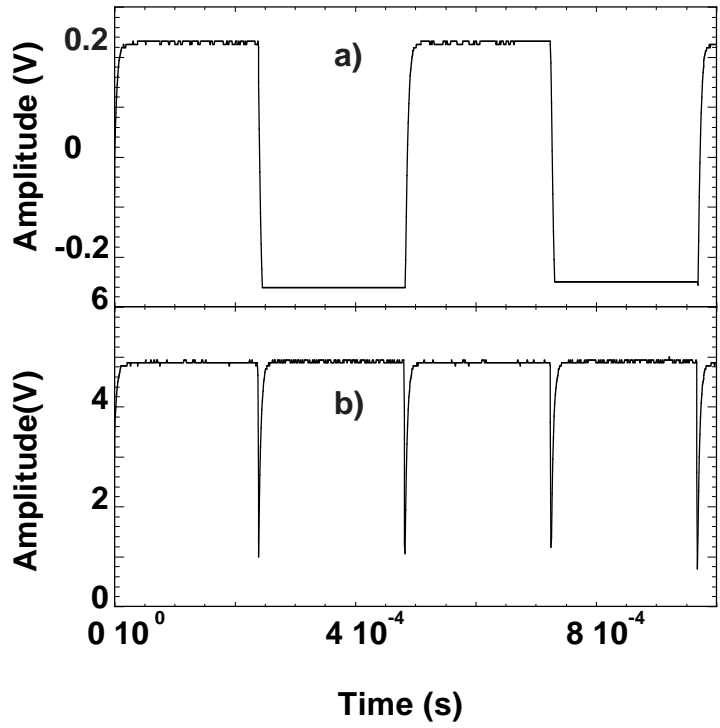

Figure 5

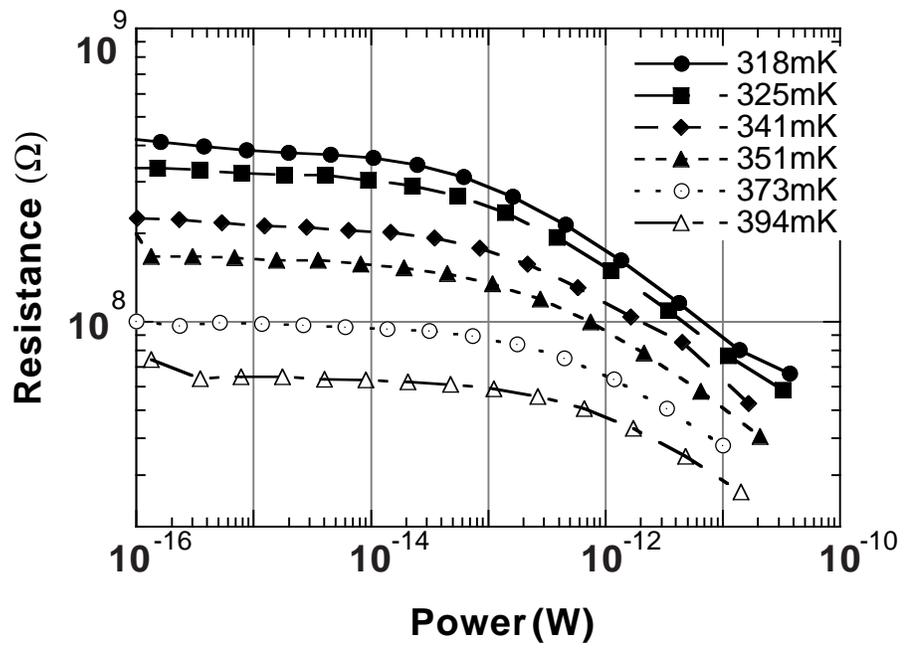

**Figure 6**

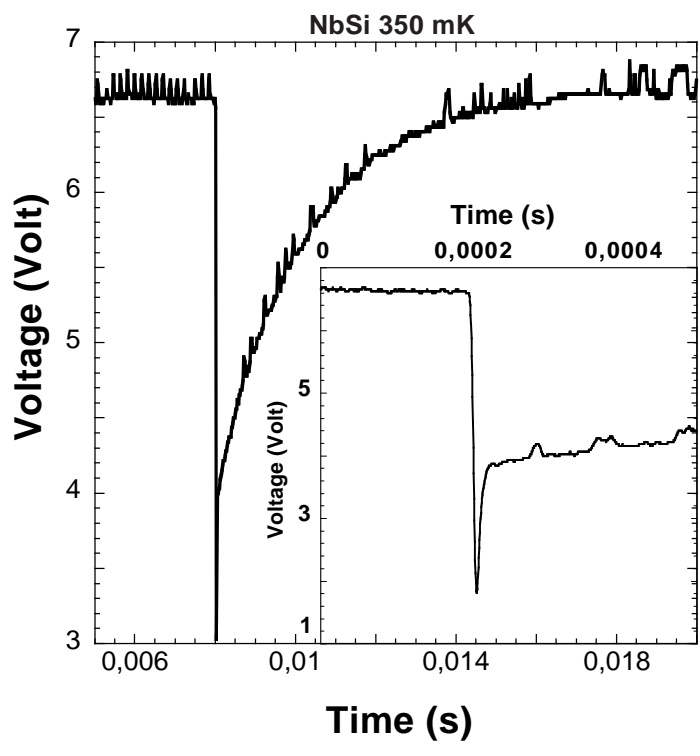

**Figure 7**